# Deep brain fluorescence imaging with minimally invasive ultra-thin optical fibers

Shay Ohayon [1,2], Antonio M. Caravaca-Aguirre [3], Rafael Piestun [3], James J. DiCarlo [1,2]

(1) Mcgovern Institute for Brain Research, Massachusetts Institute of Technology, 43 Vassar Street Cambridge, MA 02139
(2) Department of Brain and Cognitive Sciences, Massachusetts Institute of Technology, 77 Massachusetts Avenue, Cambridge, MA 02139
(3) Department of Electrical, Computer, and Energy Engineering, University of Colorado, Boulder, Colorado, 80309, USA

**Abstract**

*A major open challenge in neuroscience is the ability to measure and perturb neural activity in vivo from well-defined neural sub-populations at cellular resolution anywhere in the brain. However, limitations posed by scattering and absorption prohibit non-invasive (surface) multiphoton approaches[1,2] for deep (>2mm) structures, while Gradient Refreactive Index (GRIN) endoscopes[2–4] are thick and cause significant damage upon insertion. Here, we demonstrate a novel microendoscope to image neural activity at arbitrary depths via an ultrathin multimode optical fiber (MMF) probe that is 5-10X thinner than commercially available microendoscopes. We demonstrate micron-scale resolution, multispectral and volumetric imaging. In contrast to previous approaches[1,5–8] we show that this method has an improved acquisition speed that is sufficient to capture rapid neuronal dynamics in-vivo in rodents expressing a genetically encoded calcium indicator. Our results emphasize the potential of this technology in neuroscience applications and open up possibilities for cellular resolution imaging in previously unreachable brain regions.*

## 1. Introduction

The main potential advantage of MMF for biological endoscopy is their thin diameter (50-150 um) which induces less damage compared to thicker probes, such as GRIN lens[4,9–11]. Fiber bending remains a major challenge to MMF imaging techniques that are based on Wavefront shaping (WFS) since fiber deformation changes how modes are coupled and invalidates precomputed transformations critical for image formation[12–14]. However, in some domains, such as neuroscience, major fiber bending may not always pose a direct problem since in many experimental designs fibers can be inserted into the brain along a straight trajectory to target specific brain regions. Furthermore, minor perturbations at the distal end of the fiber have only minor effects on the imaging capability if a proper fiber is selected[15]. Other challenges, on the other hand, need to be addressed to make this technology useful for neuroscience. First, acquisition speed needs to be sufficiently high to capture rapid neural firing events (ms scale), while maintaining sufficient spatial resolution to resolve cellular level details. Previous approaches[1,5–8] that use liquid crystal spatial light modulators (LC-SLM) have not been shown to be capable of this temporal resolution which would prohibit rapid sampling of neural signals. Second, the system needs to have high collection efficiency to capture small fluorescence changes evoked by calcium transients without causing photobleaching. To date, only fixed tissue has been successfully imaged[5,15]. Third, the system should image some distance away from the fiber tip since tissue in close proximity may not be functioning properly due to possible damage inflicted by fiber insertion.

Here, we focus on addressing these challenges and propose a novel microendoscope system that is based on a digital mirror device (DMD).

## 2. Methods

### 2.1 Generating phase modulation with a digital mirror device

The basic idea behind the Lee hologram[16] technique is to create a binary mask that creates a diffraction pattern with the desired phase modulation. Following the work of [16,17], to create a 2D spatial phase mask $\Phi(x, y)$, we first define a spatial carrier wave with frequency f and rotation $\theta$, over the mirror domain [x,y]. The mask is then defined as:

(1) $$X = \cos(\theta) x + \sin(\theta) y$$

(2) $$Mask = \frac{1}{2}(1 + \cos(2\pi X - \Phi(x, y)) > 0.5$$

This mask generates three peaks on the Fourier plane since

(3) $$\frac{1}{2}(1 + \cos(2\pi x - \Phi(x, y)) = \frac{1}{2} + \frac{1}{4}e^{2\pi j x f_0}e^{-j\Phi(x,y)} + \frac{1}{4}e^{-2\pi j x f_0}e^{j\Phi(x,y)}$$

By blocking the first two terms (DC and one diffraction order) with an iris and allowing only one diffraction order to pass through, we create a 2D phase array that is up to a constant shift from the desired $\Phi(x, y)$. In the limit, each phase $\Phi(x, y)$ cannot be represented by a single mirror on the DMD, hence multiple mirrors are grouped to represent a single desired phase. We have experimented with varying sampling (number of mirrors per phase) and found that 10 is a good compromise between the size of each block and sufficient remaining mirrors to deliver the constant reference phase (see below).

### 2.2 Transmission matrix estimation and spot generation

The main derivation of the Transmission matrix method can be found in [1]. The goal of this procedure is to estimate the complex matrix K, which describes how the complex input field $E^{in}$ is mapped to the complex output field $E^{out}$. The assumption is that this mapping can be approximated by a linear relationship:

(4) $$E^{out} = K * E^{in}$$

(where * denotes matrix multiplication).
In our case, the input field $E^{in}$ can be described as a 2D array of wavelets (64x64, or 4096x1), where each wavelet can be phase shifted independently using the Lee hologram presented on the DMD. In other words, $E^{in} = e^{i\varphi(x,y)}$, where $\varphi(x,y) \in [0, \pi]$. The dimensionality of the output is determined by the resolution of the camera that images the tip of the fiber. In our case, $\dim(E^{out}) = 128x128$.

If we present many inputs, and measure many outputs of this linear system, we can recover K using the simple relationship:
(5) $$K = M(B)^{-1}$$
Where B is a matrix, representing a set of input vectors (basis), stacked in a row fashion and M represents a matrix of observed complex outputs. To solve this equation, we need to measure $E^{out}$ for each of the input $E^{in}$. However, $E^{out}$ cannot be observed directly because it is complex (i.e., only the intensity $|E^{out}|^2$ can be measured.

To recover the complex value of $E^{out}$ we use an interferometric approach. We present both $E^{in}$ and a second reference beam $r$ (i.e., a complex scalar). In our case, the second beam is actually the region surrounding $E^{in}$, which we give a fixed constant phase value (i.e., 0). By phase shifting the entire input field $E^{in}$ by a fixed phase offset one can recover $E^{out}$ (up to the an unknown fixed phase). Let's look at an example where a specific input field E is presented:

(6) $$\begin{aligned} I^\alpha = |E^{out}|^2 &= |r + e^{i\alpha}KE|^2 \\ &= |r|^2 + |e^{i\alpha}KE|^2 + 2\mathcal{R}e(\bar{r}e^{i\alpha}KE) \\ &= |r|^2 + |KE|^2 + 2\mathcal{R}e(\bar{r}e^{i\alpha}KE) \end{aligned}$$

$I^\alpha$ represents out measured image, when the input field has been phase shifted by $\alpha$. Notice that the first and second terms do not depend on $\alpha$. Remember that our goal is to recover K. Let us define $P \triangleq \bar{r}KE$ and look at the following expression:
(7) $$Q \triangleq \frac{1}{2}\left[I^0 - I^{\frac{\pi}{2}} + i\left(I^{\frac{\pi}{2}} - I^\pi\right)\right]$$
We will see that we can recover Z from Q:

(8) $$Q = [\mathcal{R}e(P) - \mathcal{R}e(iP) + i(\mathcal{R}e(iP) - \mathcal{R}e(-P))$$
Let's look at a specific index in P, and denote $P_m \triangleq a + ib$. Then equation (8) becomes:
$$Q = a + b + i(a - b)$$
Thus, $P_m = \frac{1}{2}(Re(Q) + Im(Q)) + \frac{1}{2}i(Re(Q) - Im(Q))$
A single presentation of an input field $E^j$ (4096x1) gives us (16384x1) constraints on K. To fully recover K (16384x4096), we need a full basis of inputs. The matrix derivation can be formulated as follows:

(8) $$Z = diag(r)KE$$
where Z is a complex matrix calculated from the measured intensities, K is the unknown transmission matrix and E is a set of complex field inputs ($E = [E^1, E^2, ..., E^{4096}]$ of our choice. If we let $E = I$ (the identify matrix), the solution is trivial ($K \sim Z$). However, this is not efficient in terms of SNR because only a tiny fraction of the mirrors would be turned on for any given input. Any choice of E is reasonable as long as it is orthogonal. In practice, we use the Hadamard-Walsh basis because it is balanced in terms of the mirrors that are set ON and OFF and also because it is easily invertible $H^{-1} = H^T$. We first map the Hadamard basis from the [-1,1] domain to the complex phase $[e^{i*0}, e^{i*\pi},]$ (i.e., a complex field that we can generate with our Lee-hologram). We denote the complex phase basis as H. K is easily recovered by $K \sim Z\bar{H}$. Under the assumption that K is unitary ($KK^* = I$), the inverse operation becomes:

(9) $$K^{-1} = HZ^*$$
, where * denotes the complex conjugate operator. To generate a spot at location (x,y), K is inverted and multiplied by a delta function at the desired pixel location $\delta_{index}$, where index is the flattened position of (x,y). Any pattern (with certain limitations) can be generated by multiplying it with $K^{-1}$. For example, it is possible to generate larger excitation spots which are useful for rapid sub-sampling scanning or structured illumination to target specific structures. Only the phase angle of the above derivation is needed since only phase modulation is applied.

The actual calculations are sped up considerably with a custom C++ implementation that uses Cublas (Cuda optimized matrix library for GPU). A total of 12288 patterns need to be displayed (4096x3) on the DMD to measure K. With a reasonably fast CMOS camera (MQ013MG-ON, Ximea) it takes ~8 seconds (1500Hz) to acquire those those patterns. The whole procedure takes less than 15 seconds, to perform the remaining matrix multiplication and hologram generation

### 2.3 Multispectral imaging
To allow multispectral imaging we first recognized that the DMD will diffract the two excitation beams to different directions. These diffraction patterns can be steered by controlling the carrier wave frequency and rotation such that only one excitation wavelength passes through the iris with minimal interference from the other laser. Two calibrations (one for each laser source) are needed. Rapid transitioning between the two sources is then achieved by setting the DMD with the desired pattern. In other words, we can switch between blue and green laser excitation at 22kHz.

### 2.4 Enhancement Factor Metric
To evaluate the quality of our generated spots we devised an automatic procedure that calculates the TM, generates several hundred spots and image each one of them under several neutral density filters. The multiple exposures of the same spot are then combined to form a high dynamic range image in the following way. First, for a given exposure under neutral density n, a dark image is subtracted and all pixels that are over-exposed or underexposed are removed. Then, the high dynamic range (HDR) image is calculated as

(10) $$HDR(x,y) = max_n(I(x,y,n)10^n)$$

where n corresponds to the neutral density value. The peak of the HDR image is found ($I_{peak}$), and a small neighborhood (5x5 pixels) around that region is removed. Then, the average signal inside the fiber core is calculated ($I_{avg}$). The enhancement factor is defined as $I_{peak}/I_{avg}$. We believe this approach is less susceptible to numerical instabilities from $I_{avg}$ underexposure, which can artificially inflate enhancement values. This is typically the case since the peak is several orders of magnitude brighter than the background, which exceeds most camera's dynamic range.

### 2.5 Point Spread Function Estimation
To measure the (x,y) spread of the point spread function we imaged fluorescence microspheres and fitted each individually identified blob with a 2D Gaussian by fitting (a dark subtracted image) with the following function:

(11) $$F(x,y) = Ae^{-\left[\frac{(x-x_0)^2}{2\sigma_x^2} + \frac{(y-y_0)^2}{2\sigma_y^2}\right]}$$

The FWHM of the Guassian was defined as $2\sigma\sqrt{2ln2}$, where $\sigma = \sqrt{\sigma_x^2 + \sigma_y^2}$

The signal to noise ratio (SNR) was defined as A/N, where A represents the amplitude of the 2D Gaussian and N is the average intensity measured in a background region not containing microspheres.

### 2.6 Microspheres to background separability
We used the sensitivity index (d') from signal detection theory to quantify the Z-section with the sharpest focus. d' is defined as $d' \triangleq \frac{\mu_A - \mu_B}{\sqrt{\frac{1}{2}(\sigma_A^2 + \sigma_B^2)}}$, where $\mu_A$ is the signal mean intensity, $\mu_B$ is the noise mean intensity, $\sigma_A$ is the signal standard deviation and $\sigma_B$ is the noise standard deviation.

### 2.7 Software, Data collection and processing
The software to operate the fiberscope was mostly written in Matlab (mathworks). Some modules were implemented in C++ and Compute Unified Device Architecture (CUDA) to speed up calculations. In an earlier version of the microscope we digitized PMT emissions using a DAQ capable of sampling at 500kHz. For each DMD flip, 10 samples were collected. In a newer version we developed a custom FPGA code running on a Xilinx chip that was triggered by a DMD flip, waited for 15s to allow mirror stabilization on the DMD and then generated TTLs at 8MS/s to drive the DAQ to sample emissions from the PMTs (total dwell time: 50 usecs at 20 kHz). This resulted in 180 samples / pixel, which were averaged to get the raw image data. We resampled movies such that each pixel was sampled at time t (to correct the small deviation caused by in-frame raster scan time). Unless otherwise stated, images were notch filtered to remove unwanted frequency bands (typically, 60Hz), then spatially smoothed with a Gaussian (=1.5-2m). For functional imaging we also temporally smoothed the sequence with a gaussian (=1.5-2 frames). In some plots, histogram equalization was performed on images (dynamic range stretching) to improve visualization of faint structures and to map a higher dynamic range to 8-bit images.

### 2.8 Equipment

| Item | Model | Manufacturer |
|---|---|---|
| DMD | V7000 | Vialux |
| PMT1 | H7422-40 | Hamamatsu |
| PMT1 Amp | C9999+ M9012 | Hamamatsu |
| PMT 2 | PMTSS | Thorlabs |
| PMT2 Amp | TIA60 | Thorlabs |
| DAQ | USB-2020 | Measurement Computing |
| FPGA | Mojo V3 | Embedded Micro |
| GPU | GTX 970 | NVIDIA |
| CMOS | GS3-U3-23S6M | Point Grey (FLIR) |
| 473nm Laser | MLL-FN-473-300 | Ultralaser |
| 532nm Laser: | MGL-III-532-100 | Ultralaser |
| CPU | i7-4770K | Intel |
| Objectives | 40X Plan Achromatic 0.65 NA, 10X Plan Achromatic 0.25 NA | Olympus |
| Filters | LF405/488/532/635-A-000, BrightLine® full-multiband laser filter set | Semrock |
| Fiber | 20um, 0.66NA | Neuronexus |
| Microspheres | 0.96um FS03F, 15um FS07F | Banglabs |

### 2.9 Surgery and animal handling
All procedures have been approved by the Massachusetts Institute of Technology Institutional Animal Care and Use Committee and conform to NIH standards.
Wild type mice (C56BL) from Jackson Labs were anesthetized with isoflurane (1-2%). A small borehole

(~300-500um) was drilled and 50-100nl of AAV9-hSyn-GCaMP6s was pressure injected using a custom pump (25nl/minute) to deliver the GCaMP gene. We allowed 2-3 weeks of recovery and virus expression prior to imaging. Mice were then anesthetized again and a slightly larger craniotomy was performed to expose the brain.

We used a thin wall hypodermic needle (28G) as reinforcement on the upper part of the fiber, leaving ~2mm of fiber exposed for insertion (the needle was not pushed in the brain).

Animals were then raised on a custom movable translational stage in small increments until the fiber punctured through the dura and entered the brain. At the end of the imaging session, animals were euthanized with a lethal dose of pentobarbital.

## 3. Results

The microscope design (figure 1) is based on wave front shaping (WFS)[7,18] : the light wavefront is modulated before it is coupled at the proximal side such that when it exits at the distal tip, it creates a micron-scale excitation spot at a desired (x,y) position (figure 2) and distance (z) from the fiber tip (figure 3). Sections (or the entire volume) are reconstructed by scanning the excitation spot and measuring the fluorescence emissions collected by the same fiber. To generate wave front phase modulation we use a computer-generated hologram method[14,16,17] on a digital mirror device (DMD), that allows kilohertz phase modulation (methods). Random access scanning is possible since forming a spot at (x,y,z) corresponds to programming the DMD with the pre-computed pattern. Structured illumination is also possible, allowing to dynamically changing the size of the PSF (methods). The 2D array of adjustable phases mix with a fixed reference beam in the fiber (figure 1) to generate the observed complex interference pattern at the distal tip (speckle pattern). The input-output transformation of the fiber is characterized using the transmission matrix (TM) algorithm[1] (methods) and the complex field is measured using a fast (three step) phase stepping algorithm. The entire procedure (termed "calibration") of measuring the TM and computing the holograms needed to scan the entire FOV (100x100 um) takes ~20 seconds using a fast CMOS camera and GPU optimized software (methods).

To assess the optical quality of the microendoscope we first developed a robust metric to measure the enhancement factor (EF), quantifying the strength of the excitation spot at the distal tip of the fiber (methods). We image generated spots with multiple neutral density filters and combine them to form a high dynamic range image (figure 4a) from which the EF can be calculated (methods). Typically obtained EF values were 305±125(range: 50-550) and decreased as a function of radial distance (figure 4b). To further assess the optical quality we measured the point spread function (PSF) along the emission path by imaging 0.96um fluorescent microspheres (figure 4c). The FWHM of a 2D Gaussian fitted to the blobs was 2.10±0.25 um (mean, std), with estimated SNR of 1.39±0.1(methods).

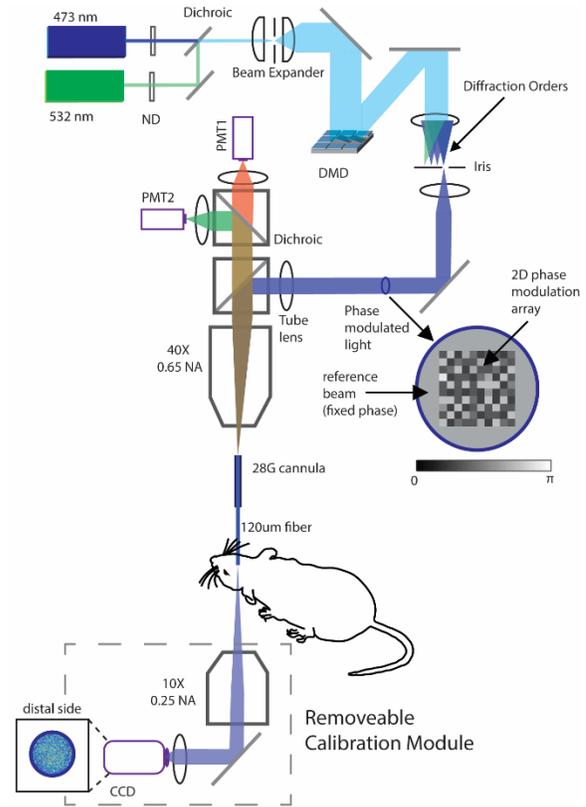

**Figure 1**. *Laser sources are combined and expanded to cover the surface of a DMD, programmed to display a Lee Hologram. Only one diffraction order is permitted at the Fourier plane. Generated phases are wrapped by a fixed phase offset serving as a reference beam. Light is focused on the proximal side of a 120um multimode fiber. Phases are scrambled in the fiber generating the interference pattern observed at the distal tip via a 10X objective and a CMOS Camera. Fluorescence emissions collected with the same fiber, split with a second dichroic and measured with two PMTs.*

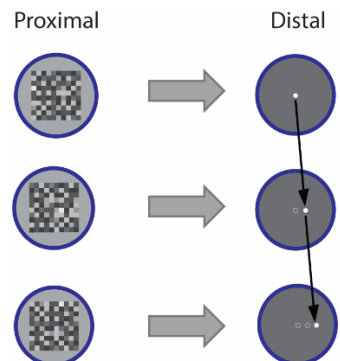

**Figure 2**. *Excitation spot is raster scanned by programming the DMD with the computed holograms. Random access scanning is possible.*

.

focus was obtained at that Z-section (figure 5). Our approach can also be used to rapidly switch between laser sources (methods) to allow multispectral imaging. To demonstrate this capability we imaged two types of microspheres with different excitation and emission spectra (figure 6).

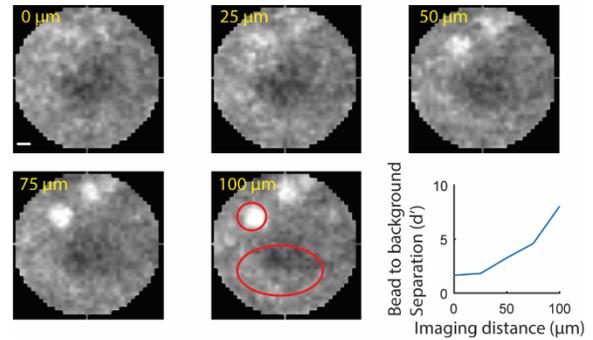

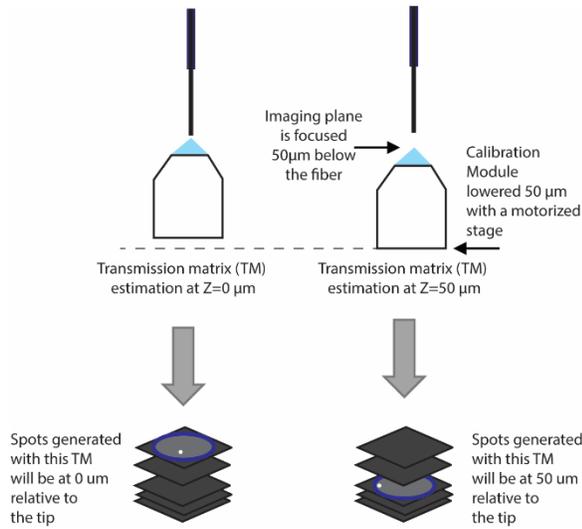

**Figure 3**. *Transmission matrix measurements taken at slightly shifted calibration camera position yield phase information to generate spots away from the fiber tip, enabling volumetric scanning.*

**Figure 5**. **d)** *Volumetric imaging of 15um microspheres in a scattered medium (2% intralipid in agarose) and a quantification of separation (d') relative to background. Regions highlighted in red used for d' calculation. Scale bar: 10 um.*

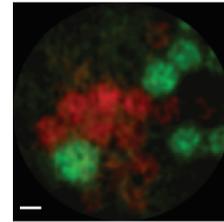

**Figure 6.** *Example of multispectral imaging of microspheres with two emission bands. Scale bar: 10 um.*

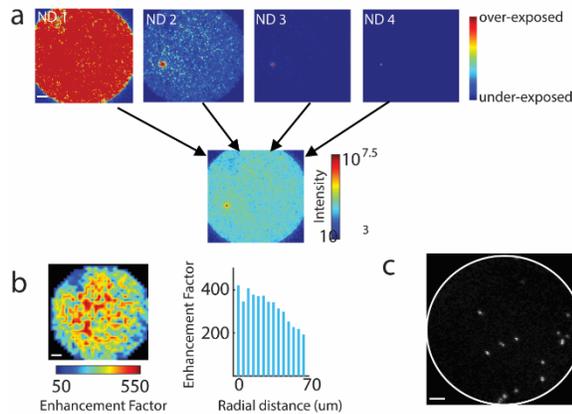

**Figure 4. a)** *Four images taken at different neutral density filters are combined to form a single high dynamic range (HDR) image from which the EF of the generated spot can be calculated.* **b)** *Spatial distribution of EF values for spots generated across the fiber core.* **c)** *Image of 0.96um fluorescence microspheres on a glass slide used to assess in-plane resolution.*

Fluorescent microspheres can be 10-1000 brighter than in-vivo biological fluorescence. To test whether our microendoscope has the sensitivity needed to capture enough photons from cells expressing GFP we set up an experiment to simultaneously image live cells on a glass slide with both the fiber from the top and a wide field epifluorescence microscope from the bottom. We imaged live baby hamster kidney (BHK) cells expressing GFP (figure 7) and found that thin fiberblast features could be easily recognized.

The ability to image away from the tip in the brain is important since cells nearest to the fiber are more likely to be damaged. To assess our ability to form excitation spots away from the fiber tip in a scattering medium we imaged 15um microspheres embedded in 2% intralipid agarose (figure 5). By shifting the calibration camera by a fixed offset and remeasuring the TM, it is possible to find the phases needed to raster scan at different planes away from the tip[19], essentially acquiring a volume without the need to physically move the fiber during imaging. We quantified the separability between microspheres embedded in this scattering medium and the background using d' (methods) and found the highest d' at 100um, suggesting the sharpest

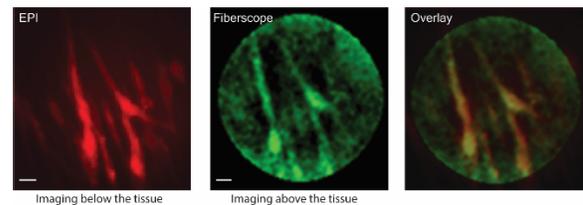

**Figure 7**. *In-vitro imaging of BHK cells expressing GFP. Comparison between full field epi imaging with a regular camera and the fiberscope.*

The main innovation of our approach is the ability to rapidly scan the entire field of view in 3D to allow real-time measurements of calcium events evoked by spikes in-vivo. We can currently achieve full frame (100x100um) acquisition speeds of ~7-10 Hz, while smaller fields of view (~ 10x10um) can be imaged at speeds exceeding 200 Hz. To assess whether the microendoscope has collection efficiency sufficient to capture small fluorescence changes at such rates, we imaged a thin hippocampal neuronal tissue culture expressing the genetically encoded calcium indicator GCaMP6f[20]. Multiple neurons could be identified in the FOV (figure 8). The fluorescence time course measured with the fiber was in good agreement with epi microscope measurements (mean correlation : 0.54, n=6).

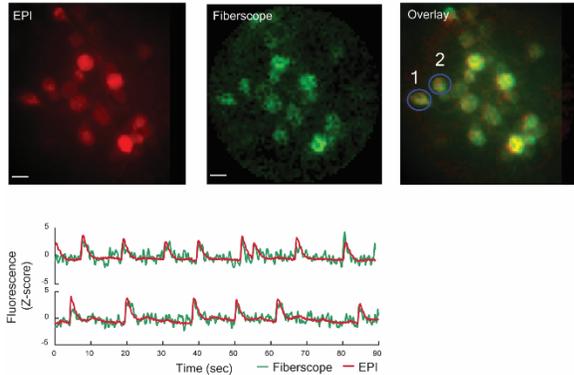

**Figure 8.** *Direct comparison between fiber imaging and epifluorescence wide field imaging of hippocampal neuronal tissue culture expressing GCaMP6f. Time-course of two ROIs highlighted showing spontaneous firing rate. Scale bar :10 um.*

These results suggest our method can be applied to record activity in-vivo. To test this directly we targeted primary visual cortex in wild-type mice expressing GCaMP, delivered via viral-mediated transduction. We slowly raised animals on a motorized translational stage towards the fiber in steps of ~50um. Upon fiber insertion, several neurons could be identified as they entered the field of view (figure 9a). Since neural activity under anesthesia is low, clearer images were obtained by calculating the standard deviation over time (figure 9b). Neurons exhibited varying degree of activity and were not significantly photobleached over several minutes exposure at ~60nW (figure 9c).

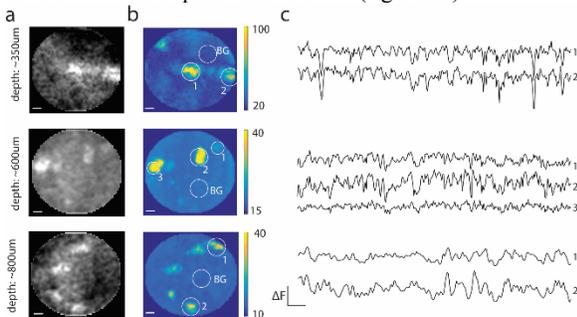

**Figure 9**. *In-vivo functional calcium imaging in wild type mouse expressing a genetically encoded calcium indicator GCaMP6s.* **a)** *Time-averaged fluorescence images. Image were taken from the same animal at multiple depths spanning ~1mm relative to the cortical surface. Bright spots correspond to neurons. Scale bar: 10 um.* **b)** *Standard-deviation of fluorescence over time showing regions that were more functionally active compared to the background. Scale bar: 10 um.* **c)** *Time course of highlighted region in (c) after background subtraction (denoted BG). Scalebar: dF: 100 (AU), 2.5 sec.*

## Conclusion

In summary, we present a new microendoscope imaging system that is based on WFS using a DMD to control light in a single fiber. To the best of our knowledge, this is the first demonstration of using multimode fibers to image live cells and in-vivo neurons and represents a significant improvement over single pixel imaging fiber photometry[21]. Although the EF in our imaging system may not be as high as one that can be obtained with LC-SLM based approaches, we demonstrate it is sufficient to acquire images which allow identification of fine features at the neuronal scale. A key improvement over previously proposed methods [22–25], which is required for live imaging, is the ability to rapidly scan a large field of view, while maintaining sufficient spatial resolution to resolve cellular details and a high enough sensitivity to precisely measure small fluorescence fluctuations. The ability to rapidly switch between two light sources and to form structured light patterns may be useful to manipulate specific cells within the field of view using optogenetics. We expect this technology will open new avenues for addressing circuit level questions in deep brain regions that have thus far been unreachable with existing imaging modalities.

**Acknowledgements**

We would like to thank Ian Wickersham for providing samples of BHK cells, Or Shemesh and Ed Boyden for providing the GCaMP6 neuronal tissue culture and discussions throughout the project. Mcgovern Institute for Brain Research for internal seed money used to fund the project. Life Sciences Research Foundation and Howard Hughes Medical Institute for providing financial support for S.O. through his postdoc. We thankfully acknowledge support from the National Institute of Health Award REY026436A.

**References**

1. Popoff, S. M. *et al.* Measuring the transmission matrix in optics: an approach to the study and control of light propagation in disordered media. *Phys. Rev. Lett.* **104,** 100601 (2010).
2. Horton, N. G. *et al.* In vivo three-photon microscopy of subcortical structures within an intact mouse brain. *Nat. Photonics* **7,** (2013).
3. Ziv, Y. & Ghosh, K. K. Miniature microscopes for large-scale imaging of neuronal activity in freely behaving rodents. *Curr. Opin. Neurobiol.* **32,** 141–147 (2015).

4. Ghosh, K. K. *et al.* Miniaturized integration of a fluorescence microscope. *Nat. Methods* **8,** 871–878 (2011).
5. Papadopoulos, I. N., Farahi, S., Moser, C. & Psaltis, D. Focusing and scanning light through a multimode optical fiber using digital phase conjugation. *Opt. Express* **20,** 10583–10590 (2012).
6. Psaltis, D., Papadopoulos, I., Farahi, S. & Moser, C. Imaging Using Multi-Mode Fibers. in *Optics in the Life Sciences* (2013). doi:10.1364/boda.2013.bm4a.1
7. Cizmár, T. & Dholakia, K. Exploiting multimode waveguides for pure fibre-based imaging. *Nat. Commun.* **3,** 1027 (2012).
8. Čižmár, T. & Dholakia, K. Shaping the light transmission through a multimode optical fibre: complex transformation analysis and applications in biophotonics. *Opt. Express* **19,** 18871–18884 (2011).
9. Bocarsly, M. E. *et al.* Minimally invasive microendoscopy system for in vivo functional imaging of deep nuclei in the mouse brain. *Biomed. Opt. Express* **6,** 4546–4556 (2015).
10. Szabo, V., Ventalon, C., De Sars, V., Bradley, J. & Emiliani, V. Spatially selective holographic photoactivation and functional fluorescence imaging in freely behaving mice with a fiberscope. *Neuron* **84,** 1157–1169 (2014).
11. Mekhail, S. P., Arbuthnott, G. & Chormaic, S. N. Advances in Fibre Microendoscopy for Neuronal Imaging. *Optical Data Processing and Storage* **2,** (2016).
12. Plöschner, M., Tyc, T. & Čižmár, T. Seeing through chaos in multimode fibres. *Nat. Photonics* **9,** 529–535 (2015).
13. Farahi, S., Ziegler, D., Papadopoulos, I. N., Psaltis, D. & Moser, C. Dynamic bending compensation while focusing through a multimode fiber. *Opt. Express* **21,** 22504–22514 (2013).
14. Caravaca-Aguirre, A. M., Niv, E., Conkey, D. B. & Piestun, R. Real-time resilient focusing through a bending multimode fiber. *Opt. Express* **21,** 12881–12887 (2013).
15. Caravaca-Aguirre, A. M. & Piestun, R. Single multimode fiber endoscope. *Opt. Express* **25,** 1656 (2017).
16. Lee, W.-H. *Computer-generated Holograms: Techniques and Applications*. (1978).
17. Conkey, D. B., Caravaca-Aguirre, A. M. & Piestun, R. High-speed scattering medium characterization with application to focusing light through turbid media. *Opt. Express* **20,** 1733–1740 (2012).
18. Vellekoop, I. M., Lagendijk, A. & Mosk, A. P. Exploiting disorder for perfect focusing. *Nat. Photonics* (2010). doi:10.1038/nphoton.2010.3
19. Mahalati, R. N., Gu, R. Y. & Kahn, J. M. Resolution limits for imaging through multi-mode fiber. *Opt. Express* **21,** 1656–1668 (2013).
20. Chen, T.-W. *et al.* Ultrasensitive fluorescent proteins for imaging neuronal activity. *Nature* **499,** 295–300 (2013).
21. Gunaydin, L. A. *et al.* Natural neural projection dynamics underlying social behavior. *Cell* **157,** 1535–1551 (2014).
22. Barankov, R. & Mertz, J. High-throughput imaging of self-luminous objects through a single optical fibre. *Nat. Commun.* **5,** 5581 (2014).
23. Kolenderska, S. M., Katz, O., Fink, M. & Gigan, S. Scanning-free imaging through a single fiber by random spatio-spectral encoding. *Opt. Lett.* **40,** 534–537 (2015).
24. Choi, Y. *et al.* Scanner-free and wide-field endoscopic imaging by using a single multimode optical fiber. *Phys. Rev. Lett.* **109,** 203901 (2012).
25. Kim, D. *et al.* Toward a miniature endomicroscope: pixelation-free and diffraction-limited imaging through a fiber bundle. *Opt. Lett.* **39,** 1921–1924 (2014).